\documentclass{aa}
\usepackage{natbib,twoopt}
\bibpunct{(}{)}{;}{a}{}{,}  % to follow the A&A style
\usepackage[varg]{txfonts}
\usepackage{graphicx}

\def\Mpg{{m$_\mathrm{pg}$}}
\def\Ubes{{U$_\mathrm{b}$}}
\def\Bbes{{B$_\mathrm{b}$}}
\def\Vbes{{V$_\mathrm{b}$}}
\def\ustr{{u$_\mathrm{s}$}}
\def\vstr{{v$_\mathrm{s}$}}
\def\bstr{{b$_\mathrm{s}$}}
\def\ystr{{y$_\mathrm{s}$}}
\def\c1str{{$\mathrm{c_1}$}}
\def\hbw{{Hbw}}
\def\hbn{{Hbn}}
\def\Hbet{{H$_\beta$}}
\def\Hgam{{H$_\gamma$}}
\def\Hdel{{H$_\delta$}}
\def\Heps{{H$_\epsilon$}}
\def\Hzet{{H$_\zeta$}}
\def\KCa{{\ion{Ca}{II}\,K}}
\def\lII{{l$^\mathrm{II}$}}
\def\bII{{b$^\mathrm{II}$}}
\def\Teff{{T$_\mathrm{eff}$}}
\def\MB{{M$_\mathrm{B}$}}
\def\AV{{A$_\mathrm{V}$}}
\def\AB{{A$_\mathrm{B}$}}

\begin{document}

\title{A catalog of early-type stars toward the Galactic center\thanks{Based
    on observations collected at the European Southern Observatory, La~Silla,
    Chile (ESO programme 59.D-0143, 271.B-5007) } \thanks{The catalog of
    early-type stars is only available in electronic form at the CDS via
    anonymous ftp to cdsarc.u-strasbg.fr (130.79.128.5) or via
    http://cdsweb.u-strasbg.fr/cgi-bin/qcat?J/A+A/}}

\author{P.~Grosb{\o}l}
\offprints{P.~Grosb{\o}l, \email{pgrosbol@eso.org}}
\institute{
        European Southern Observatory,
        Karl-Schwarzschild-Str.~2, D-85748 Garching, Germany
}
\date{Received ???; accepted ???}

%----------------------------------Abstract-------------------------
\abstract{It is still unclear whether the Sagittarius spiral arm is a major
  spiral arm in the Galaxy or whether it just outlines a region of enhanced
  star formation because of the local compression of gas.  The best way to
  separate these scenarios out is to study the kinematics across the arm to
  determine the velocity perturbation it induces. }
{A survey of early-type stars in the direction of the Galactic center is
  performed covering an area of 100 square degrees with the aim of identifying
  candidates for a radial velocity study.}
{Objective prism plates were obtained with the 4\degr\ prism on the ESO
  Schmidt telescope using IIaO, 4415, and IIIaJ emulsions.  The plates were
  digitized and more than 100k spectra were extracted down to a limiting
  magnitude of B = 15$^\mathrm{m}$.  The spectra were cross-correlated with a
  template with Balmer lines, which yielded a candidate list of 12\,675
  early-type stars.  Magnitudes and equivalent widths of strong lines were
  calculated from the spectra, which allowed us to estimate the individual
  extinctions and distances for 11\,075 stars.}
{The survey identified 9\,571 candidate stars with a spectral type earlier than
  A1 and B $<$ 14\fm5 out to distances of more than 2\,kpc, which is beyond
  the Sagittarius arm. This is indicated by the increase of absorption in the
  plane at distances larger than 0.5\,kpc. }{}
\maketitle

%-----------------------------Section: Introduction-----------------
\section{Introduction}
It is well known that young objects such as OB associations and
\ion{H}{II}-regions outline a four-armed spiral structure in the Galaxy
\citep{georgelin76, russeil07, vallee14, reid14}.  The dynamical nature of
these arms is much less certain.  In the case of the Sagittarius arm, two main
scenarios are possible: that it is either i) a major spiral arm in the Galaxy
with a significant mass \citep{drimmel00} associated with a spiral density
wave \citep{lin64}, or ii) a secondary gas compression, with increased star
formation but little additional mass\citep{yanez08, englmaier99}.  Whereas a
direct estimate of the radial surface-density variation can be performed
toward the Perseus arm \citep{monguio15}, a direct measurement of the stellar
densities in the Sagittarius arm is very difficult owing to the heavy
attenuation by dust toward the Galactic center.  Another option for detecting
a density enhancement in the arm is to observe the velocity variations of
stars across it.  The detection of a systematic radial velocity variation
would strongly suggest the presence of a significant mass associated with the
arm.  However, the stellar population used for this type of study must be old
enough to allow the stars to respond to a density perturbation
\citep{wielen77}.  On the other hand, it should not be so old that its
velocity dispersion has increased significantly since this would reduce
its response and make a detection more difficult \citep{lin69}.  An age range
of 0.1--1\,Gyr, corresponding to late B stars to early A stars, fulfills these
criteria \citep{dehnen98}.  From a kinematic point of view, the direction
toward the Galactic center has the advantage of minimizing the influence of
galactic rotation even though crowding is an issue.

The first step of such a kinematic study is to establish a reliable candidate
list of early-type stars (e.g., B and A stars) toward the Galactic center. The
high absorption by dust makes it impossible to obtain a reliable list using
broadband visual colors.  Although intermediate-band $uvby\beta$ photometry
\citep{stromgren66} could be used, the lack of wide-field instruments with
these filters in the Southern hemisphere excludes this option.  Even
near-infrared (NIR) wide-field surveys like 2MASS \citep{2mass} or VVV
\citep{vvv,saito12}, which are much less affected by extinction, cannot safely
distinguish highly reddened early-type stars from late-type giants.  An
alternative is to use objective prism exposures with Schmidt telescopes,
although overlapping spectra due to crowding is a concern.  The presence of
Balmer lines clearly identifies stars with spectral class F or earlier, even
on low dispersion spectra.  Furthermore, the strength of Balmer and
\ion{Ca}{II} lines allow us to break the ambiguity in estimating extinctions
from stellar colors \citep{payne29}.

The paper presents a search for early-type stars in the direction toward the
Galactic center using objective prism plates.  The next section describes the
observations and basic reductions of the spectra, including astrometric
calibrations.  Synthetic photometry derived from the spectra and the fits to
Balmer lines, if present, are presented in Sect.~\ref{phot}, whereas
extinctions and distances for early-type stars are provided in
Sect.~\ref{extinction}.  A comparison with major NIR surveys is given in
Sect.~\ref{2mass}.  Finally, the spacial distribution of the early-type stars
is shown in Sect.~\ref{distribution} while a general discussion of the sample
and conclusions are provided in Sect.~\ref{discussion}.

%------------------Section: Observations and reductions--------------
\section{Observations and reductions}
All objective prism plates were obtained using the ESO 1m Schmidt telescope at
La Silla with a scale of 67.5\arcsec~mm$^{-1}$.  Its $4\degr$ prism was used,
providing a nominal dispersion of 45~nm~mm$^{-1}$ at 434~nm.  A first set of
four plates were taken in 1977, with an exposure time of 8~min without filter
using a IIa-O emulsion.  The individual plates covered a
5\degr$\times$5\degr\ field.  The spectra were widened to 0.2~mm on the plates
which had a seeing in the range of 2-3\arcsec.  These plates were not used in
the final analysis because of their coarse-grained emulsion, limited spectra
range, and low sensitivity.

To overcome these issues, a second set of plates were taken at the ESO Schmidt
in 1997, shortly before the telescope was de-commissioned, using sensitized
4415 films and IIIa-J plates. The broad spectral sensitivity of these
emulsions resulted in longer spectra which were not widened to reduce problems
with overlaps.  Both emulsions were developed in D-19 for 5$^\mathrm{m}$ at
20\degr~C. A full list of plates is given in Table~\ref{tab:plates}.
\begin{table*}
\caption[]{List of Schmidt plates used.}
\label{tab:plates}
\centering
\begin{tabular}{lccccccl}
 \hline\hline
 Plate & RA$_{2000}$ & DEC$_{2000}$ & Start time & Exposure & Emulsion &
    Seeing & Remarks \\ \hline
 P02310 & 17:35:26.3 & -31:25:54 & 1977-09-16 18:35 &
    8$^\mathrm{m}$ &  IIa-O & 2-3\arcsec & Spectra widened \\ 
 P02311 & 17:35:24.8 & -26:26:54 & 1977-09-16 19:06 &
    8$^\mathrm{m}$ &  IIa-O & 2-3\arcsec & Spectra widened \\ 
 P02312 & 17:55:26.6 & -31:25:27 & 1977-09-16 19:28 &
    8$^\mathrm{m}$ &  IIa-O & 2-3\arcsec & Spectra widened \\ 
 P02313 & 17:55:25.0 & -26:25:27 & 1977-09-16 19:50 &
    8$^\mathrm{m}$ &  IIa-O & 2-3\arcsec & Spectra widened \\ 
 P13133 & 17:46:53.7 & -29:01:50 & 1997-04-11 07:25 &
    5$^\mathrm{m}$ &  4415 & 1.0\arcsec & Film, without prism \\ 
 P13173 & 17:46:55.1 & -31:06:25\tablefootmark{a} & 1997-05-05 06:40 &
   10$^\mathrm{m}$ &  4415  & 0.9\arcsec & Film \\
 P13174 & 17:36:41.6 & -28:23:47\tablefootmark{a} & 1997-05-05 07:10 &
   10$^\mathrm{m}$ &  4415  & 0.9\arcsec & Film \\ 
 P13177 & 17:37:30.7 & -33:44:00\tablefootmark{a} & 1997-05-06 07:36 &
   10$^\mathrm{m}$ &  4415  & 1.6\arcsec & Film \\ 
 P13178 & 17:57:15.8 & -28:30:11\tablefootmark{a} & 1997-05-06 08:00 &
   10$^\mathrm{m}$ &  4415  & 1.6\arcsec & Film \\ 
 P13180 & 17:57:02.5 & -33:29:33\tablefootmark{a} & 1997-05-07 07:38 &
   10$^\mathrm{m}$ &  4415  & 0.9\arcsec & Film \\ 
 P13225 & 17:37:25.9 & -33:45:58\tablefootmark{a} & 1997-08-27 00:08 &
   10$^\mathrm{m}$ & IIIa-J & 1.7\arcsec \\ 
 P13226 & 17:46:47.4 & -31:08:27\tablefootmark{a} & 1997-08-27 00:45 &
   10$^\mathrm{m}$ & IIIa-J & 1.7\arcsec \\
 P13231 & 17:36:33.9 & -28:25:30\tablefootmark{a} & 1997-08-28 23:41 &
   10$^\mathrm{m}$ & IIIa-J & 1.7\arcsec \\
 P13232 & 17:57:08.4 & -28:31:35\tablefootmark{a} & 1997-08-28 00:05 & 
   10$^\mathrm{m}$ & IIIa-J & 1.7\arcsec & Likely a IIIa-F
   emulsion\tablefootmark{b} \\ 
 P13233 & 17:56:55.7 & -33:30:59\tablefootmark{a} & 1997-08-28 00:28 &
   10$^\mathrm{m}$ & IIIa-J & 1.8\arcsec \\ \hline
\end{tabular}
\tablefoot{ \tablefoottext{a}{Telescope coordinates.  The field is offset by
    ~2\degr\ in declination because of the prism.}  \tablefoottext{b}{Although
    the plate envelop lists the emulsion as IIIa-J, both spectral range and
    photographic contrast suggest it is a IIIa-F plate.}  }
\end{table*}

All plates were digitized on the PDS~1010 microdensitometer at ESO in density
mode.  Owing to the limited size of the PDS stage, the plates were scanned in
four parts of $\sim$16~cm$\times$16~cm, with a 1~cm overlap region.  The
plates were visually aligned so that the scan direction was along the
dispersion.  The actual alignment was measured on each scan and found to be
better than 0\fdg3 on average, except for the short, widened IIa-O spectra,
where the error was around 0\fdg5 with a maximum of 1\degr.  The
digitalization was made in a meander pattern with a speed of 3~cm~sec$^{-1}$,
which provided a correct recording of specular densities $D<3.0$ for a step
function.  The IIa-O plates were scanned with a squared aperture of 20~$\mu$m,
while 10~$\mu$m was used for the finer-grained emulsions.  With a step size
equal to the aperture, this resulted in individual scans of 8k$\times$8k and
16k$\times$16k pixels, respectively. The density was set to zero on a part of
the plate without emulsion while chemical fog, $D_{fog}$, was measured as the
smallest density along the unexposed edge of the plates.

The ESO Schmidt did not provide a sensitometry wedge on the plates which made
a direct measurement of the density-to-exposure relation impossible.  To
estimate the characteristic curve for the emulsions, five direct V-band
exposures, which were centered on the plate fields, were obtained with the WFI
CCD camera at 2.2m MPI/ESO telescope at La Silla.  Even with a total exposure
of 9~sec per field, the WFI/CCD frames were significantly deeper than the
plates and reached V=17\fm5 with errors of 0\fm01.  The WFI data were reduced
with the ESO Image Survey pipeline \citep{nonino99}.  The magnitudes of point
sources on the reduced frames were estimated using SExtractor \citep{bertin96}
and cross-identified with objects on the objective plates.  The peak
photographic density in the region between the \Hbet\ and \Hgam\ lines were
measured and plotted as a function of the V magnitudes, which yielded the
characteristic curves for the emulsions.  The formula by \citet{sampson25} was
used to compute relative intensities from densities
\begin{equation}
 \log(I/I_0) = \gamma^{-1} \log(10^{(D-D_{fog})} - 1.0),
 \label{eq:dens}
\end{equation}
where $\gamma$ is the photographic ``contrast''.  The main parameters for the
characteristic curves are given in Table~\ref{tab:gamma} ,where the number of
stars used, the contrast, and saturation density $D_s$ are listed. The inertia
point V$_\mathrm{ip}$, defined as the extrapolation of the linear part of the
characteristic curve to zero density, is also given in V-band magnitudes.

\begin{table}
\caption[]{Photographic contrast $\gamma$, inertia point V$_\mathrm{ip}$, and
  saturation density $D_s$ for emulsions. The offset $\Delta_e$ of spectral
  cutoff relative to the H$_\gamma$ line is also listed. }
\label{tab:gamma}
\begin{tabular}{lrcccc}
\hline\hline
 Emulsion &  No. &  Contrast $\gamma$ & V$_\mathrm{ip}$ & $D_s$
 & $\Delta_e$ (mm)\\ \hline 
 IIa-O    &  94  &   1.86$\pm0.06$    &     14\fm0     &  3.0 & 1.023 \\
 4415     & 623  &   2.66$\pm0.08$    &     15\fm7     &  4.0 & 2.816 \\
 IIIa-J   & 126  &   3.02$\pm0.16$    &     15\fm0     &  3.9 & 1.640 \\
 IIIa-F   &  14  &   2.07$\pm0.38$    &     17\fm4     &  4.4 & 2.901 \\ \hline
\end{tabular}
\end{table}

\begin{figure}
\resizebox{\hsize}{!}{\includegraphics{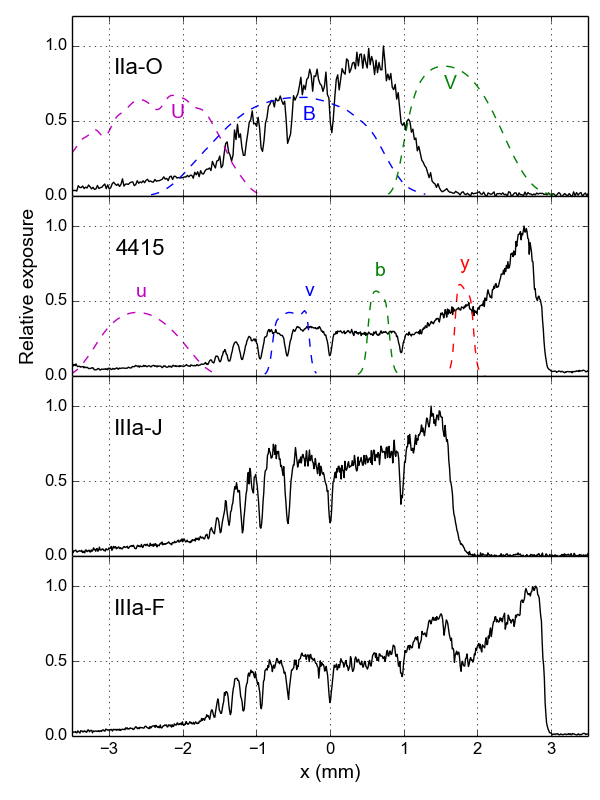}}
\caption[]{Early-type template spectra for the 4 emulsions.  The relative
  intensities of the spectra are plotted as a function of the linear distance
  from the H$_\gamma$ line in the dispersion direction. For reference,
  passbands for the Bessel filters are over-plotted on the IIa-O plate, while
  Str\"{o}mgren filters are shown on the 4415 spectrum.}
\label{fig:template}
\end{figure}

Several early-type spectra were identified visually on the plates and averaged
for each emulsion (see Fig.~\ref{fig:template}).  Preliminary positions of
candidate spectra were obtained by computing the linear correlation
coefficient between the scans and the one-dimensional (1D) template
spectra. The 4415 template was used for plate P13232 to match its spectral
range.  A selection threshold of 0.5 for the correlation coefficient gave
initial candidate lists of 41\,888, 135\,413, and 99\,027 spectra for the
emulsions IIa-O, 4415 and IIIa-J, respectively.  A search with SExtractor was
also performed and yielded similar results, with many multiple detections
along the spectra, however.

The extraction of 1D spectra from the scans was done by first analyzing a
small rectangular region around each candidate position.  The marginal
distribution perpendicular to the dispersion, taking into consideration the
tilt of the spectra measured for the scan, was computed and a search for peaks
made.  The spectrum was rejected as overlapping if several peaks were detected
near to the main peak.  The 1D spectra were extracted by converting densities
to intensities using Eq.~\ref{eq:dens}, subtracting nearby sky, and adding the
intensities across the spectrum with the weights derived from the marginal
distribution.  The emulsions have a relative sharp spectral cutof, which was
easy to identify.  The offsets between the cutoff, defined as the half
intensity point, and the \Hgam\ lines was estimated from several early-type
spectra.  The average offset $\Delta_e$ for each emulsion is given in
Table~\ref{tab:gamma}.  The spectra for which both \Hgam\ and \Hdel-lines had
the expected location and a comparable equivalence width suggest that
$\Delta_e$ could be determined with an error of 22$\mu$, 9$\mu$, and 18$\mu$
for the emulsions IIa-O, 4415 and IIIa-J, respectively.  As reference position
for the spectra, the location of the \Hgam\ line was used in the dispersion
direction while the intensity weighted position was used perpendicular to it.
For late-type spectra with no Balmer lines, the location of \Hgam\ was
estimated using the offset $\Delta_e$ derived for the emulsion type.

The dispersion relation for the plates was derived by fitting a
3$^\mathrm{rd}$ order polynomial to the six Balmer line (i.e., \Hbet\ to H9)
that were identified on spectra of early-type stars.  This gave a linear term
of 45.61 nm mm$^{-1}$ at \Hgam\ with an error of 0.027 nm.

%--------------------- Section: Astrometry -----------------------
\subsection{Astrometry}
\label{astr}
The brightest sources on each scan were visually cross-identified with the
Tycho-2 Catalogue \citep{hog00} to obtain astrometric calibrations that yield
on average 50 matches.  A linear transformation gave a standard deviation in
the range of 2-4\arcsec\ and was used to find additional matches.  An average
of 200 stars in the Tycho-2 Catalogue (varying from 42 to 463 depending on the
fields) was found to agree with the positions of the spectra within an error
of 10\arcsec.  These stars were used for a full astrometric solution, which
was done separately for right ascension and declination using five terms that
include square and cross terms of X and Y coordinates.  Pairs with residuals
larger than 2.5\arcsec\ and 4.5\arcsec\ for right ascension and declination,
respectively, were rejected leaving between 26 and 248 Tycho-2 stars per scan.
The formal standard deviations for the transformations were around
0.5\arcsec\ for the 4415 emulsion in both coordinates, whereas the other
emulsions had errors in the range 0.8-1.5\arcsec\ in declination as a result
of their more shallow sensitivity cutoff.

The actual astrometric errors are larger since the Tycho-2 stars are unevenly
distributed on the scans and the transformations were extrapolated to the
edges.  An estimate of the maximum errors can be made by comparing the
positions of stars located in the overlapping areas of different scans.  The
standard deviations of the differences for $\alpha$ and $\delta$ are given in
Table~\ref{tab:trans} and suggest that the error in right ascension is
approximately 1\arcsec,\ while declinations are determined to be better than
3\arcsec\ for the P131 plates and 6\arcsec\ for the other emulsions.
Duplicate sources (i.e., spectra of the same source occurring on several scans
or plates) were identified by matching their coordinates, magnitudes, and
spectral gradient around \Hgam.  The brightest of such duplicates was selected
as the prime spectrum.  In the case of the P132 plates, spectra on P13223 were
given the lowest priority because of the uncertainty of the plate emulsion.

\begin{figure}
\resizebox{\hsize}{!}{\includegraphics{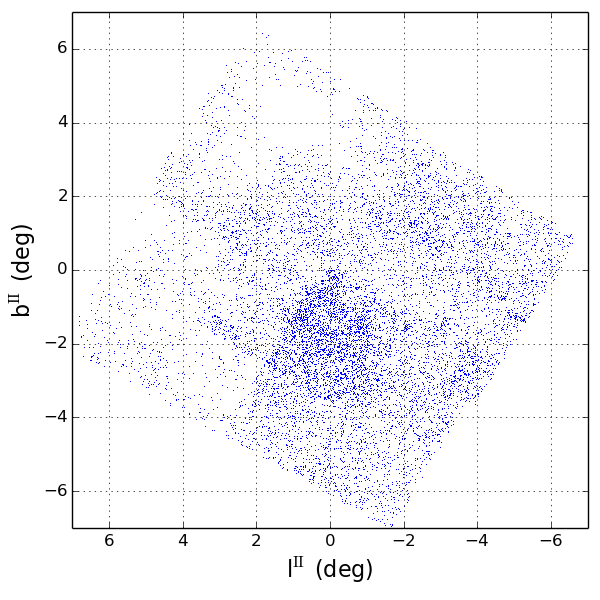}}
\caption[]{The distribution of stars with Balmer lines on the P132
  plates in galactic coordinates. }
\label{fig:lbd}
\end{figure}

The distribution of spectra with Balmer lines are shown in galactic coordinate
in Fig.~\ref{fig:lbd} where a deficiency of stars in the regions of the Pipe
Nebular around (\lII,\bII) = (1\degr,4\degr) is clearly visible.  It is also
noted that plate P13232 is significantly shallower than the other plates in
the P132 set.

%---------------------Section: Calibrations-----------
\subsection{Synthetic photometry}
\label{phot}
Synthetic photometry can be derived from the spectra since they were converted
to relative intensities and their dispersion relation was estimated.
Instrumental magnitudes were computed for several standard photometric bands
using transmission curves for the Bessel broadband and Str\"{o}mgren
intermediate-band filters from the ESO filter list\footnote[1]{see
  http://filters.ls.eso.org/} as given in Table~\ref{tab:filters} and shown in
Fig.\ref{fig:template}.  The flux in each band was computed by converting its
transmission curve into the linear scale of the objective prism spectra
relative to \Hgam, correcting for the change in bin size, and convoluting it
with the intensity along the spectra.  Only the 4415 and IIIa-F emulsions
cover the full wavelength range of all filters.  Fluxes for \Vbes\ and
\ystr\ could not be computed for the other plates.  The sensitivity cutoff of
the emulsions was the primary spectral and positional reference for the
sources.  To improve the accuracy of the location of \Hgam, a template
consisting of the Balmer lines \Hgam\ through \Hzet\ was created and
correlated with all spectra.  For sources with a correlation peak larger than
5$\sigma$, the \Hgam\ line was refitted and used as wavelength reference.

\begin{table}
\caption[]{Central wavelength, width, and ESO filter numbers for filters.}
\label{tab:filters}
\begin{tabular}{lrrl}
\hline\noalign{\smallskip}
 Filter & $\lambda_c$ (nm) & $\Delta\lambda$ (nm) & ESO filter\\
\noalign{\smallskip}\hline\noalign{\smallskip}
 \Ubes   & 354.0 &  53.7 & ESO 602 \\
 \Bbes   & 422.1 &  94.3 & ESO 603 \\
 \Vbes   & 542.1 & 104.8 & ESO 606 \\
 \ustr   & 350.3 &  31.8 & ESO 316 \\
 \vstr   & 412.4 &  19.4 & ESO 317 \\
 \bstr   & 467.9 &  16.9 & ESO 318 \\
 \ystr   & 548.6 &  21.6 & ESO 319 \\
 \hbn    & 487.9 &   2.8 & ESO 320 \\
 \hbw    & 487.9 &  15.3 & ESO 321 \\
\noalign{\smallskip} \hline
\end{tabular}
\end{table}

\begin{table}
\caption[]{Astrometric and photometric uncertainties.}
\label{tab:errors}
\begin{tabular}{l rrr rrr}
\hline\hline
 Plates &
 \multicolumn{2}{c}{P023} & \multicolumn{2}{c}{P131} &
 \multicolumn{2}{c}{P132}  \\
 quantity & \multicolumn{1}{c}{$\sigma$} & \multicolumn{1}{c}{n} &
 \multicolumn{1}{c}{$\sigma$} & \multicolumn{1}{c}{n} &
 \multicolumn{1}{c}{$\sigma$} & \multicolumn{1}{c}{n} \\
\noalign{\smallskip}\hline\noalign{\smallskip}
\Mpg          &     0.06 & 5374 &     0.14 & 8910 &     0.25 & 5613 \\
$\alpha$      & 0\farcs9 & 5374 & 0\farcs9 & 8910 & 1\farcs1 & 5613 \\
$\delta$      & 5\farcs4 & 5374 & 1\farcs6 & 8910 & 6\farcs1 & 5613 \\
B             &     0.17 &  296 &     0.25 &  285 &     0.23 &  136 \\
(u-b)         &     0.17 &   25 &     0.15 &   13 &     0.08 &    7 \\
c$_\mathrm{1}$ &     0.07 &   25 &     0.10 &   13 &     0.12 &   15 \\
H$_\beta$      &        - &    - &     0.03 &    4 &        - &    - \\
\noalign{\smallskip} \hline
\end{tabular}
\end{table}

The transformation to a standard system was done for the broadband colors
using the Catalogue of Homogeneous Means in the UBV System
\citep{mermilliod91} while the Str\"{o}mgren filters were calibrated with the
Catalogue of uvby-beta Data \citep{hauck98}.  Although almost 500 UBV and 50
uvby measurements were found in these catalogs, crowding and saturation issues
significantly limit the useful sample.  All stars in clusters and with
B<9$^\mathrm{m}$ were excluded from the calibrations. Furthermore, the spectra
were checked visually to remove stars with crowding issues.  Errors and number
of standards used are given in Table~\ref{tab:errors}, while the coefficients
of the linear transformation from standard to instrument system are listed in
Table~\ref{tab:trans}.  In addition to the synthetic photometry, Gaussian
absorption line profiles were fitted to the \Hbet (486.13nm) , \Hgam
(434.05nm), \Hdel (410.17nm), \Heps (397.01nm), and \KCa (393.37nm) lines from
which their equivalent width (EW) was estimated.  Spectra were classified as
early-type (i.e., spectral class F or earlier) if their \Hgam, \Hdel, and
\Heps\ lines were all significant at a 3$\sigma$ level and if their central
wavelengths agreed within 0.5\,nm.  This gave 1\,076, 6\,116, and 12\,675
early-type candidates for the plate sets P023, P131, and P132, respectively.

\begin{figure}
\resizebox{\hsize}{!}{\includegraphics{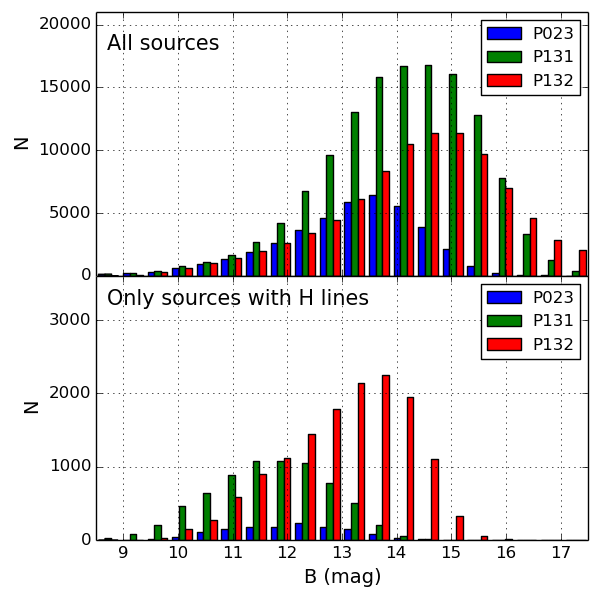}}
\caption[]{Histograms of B magnitudes of sources for the three plate sets.
  Upper panel shows all sources while the lower one includes only sources with
  Balmer lines. }
\label{fig:Bhist}
\end{figure}

The distributions of B-magnitudes for sources on the three sets of plates are
shown in Fig.~\ref{fig:Bhist}.  The P023 plates are the most shallow ones as
expected.  Although significantly more sources were found on the P131 plates
than on the P132 set, the latter is almost one magnitude deeper than the
former.  This is caused by the high relative sensitivity in the red of the
4415 emulsion (see Fig.\ref{fig:template}), which favors the detection of
nearby red dwarfs.  The number of sources in the P023 and P132 samples agrees
well above the limiting magnitude of the P023 plates which are nearly
$2^\mathrm{m}$ shallower than the P132 set.

Internal errors of magnitudes were estimated based on the pixel-to-pixel noise
along the spectra but were significantly smaller than external and systematic
errors, such as variations of the characteristic curves as a function of
wavelength, inhomogeneities across the plates, and contamination by nearby
spectra.  The limiting magnitude for the early-type sources on the P132 plates
is B=13\fm5 at a 95\% completeness level, assuming a smoothly increasing
source distribution.

%--------------------- Section: Physical parameters -----------
\section{Extinction and distances}
\label{extinction}
The first step toward deriving physical parameters for  stars is to
estimate  individual extinctions, which are expected to be significant towards the Galactic center.  This was done on the P132 plates since they
provide the faintest sample of early-type sources.  Unfortunately, the IIIa-J
emulsion does not cover the Str\"{o}mgren \ystr-band, making it impossible to
use the reddening-independent Str\"{o}mgren indices.  Also the Bessel
\Vbes-band is only partly within its sensitivity range, which leaves only
combinations of the three bluest Str\"{o}mgren filters.  The two main options
of color-color diagrams (CCD) are \c1str\ vs. (u-b) and (v-b) vs. (u-v), where
the latter was preferred due to the higher photometric error in \c1str.  The
standard problem of using these CCDs to estimate extinction is a local
maximum in these indices which makes it difficult to distinguish between
highly reddened early-type stars and less reddened late-type stars.  This
ambiguity can be resolved using the measured EWs of the Balmer and
\KCa\ lines.  The presence of Balmer lines indicates that the star is earlier
than G, while a significant \KCa\ line places it later than A0 (i.e., the local
maximum of the (u-v) index).

\begin{figure}
\resizebox{\hsize}{!}{\includegraphics{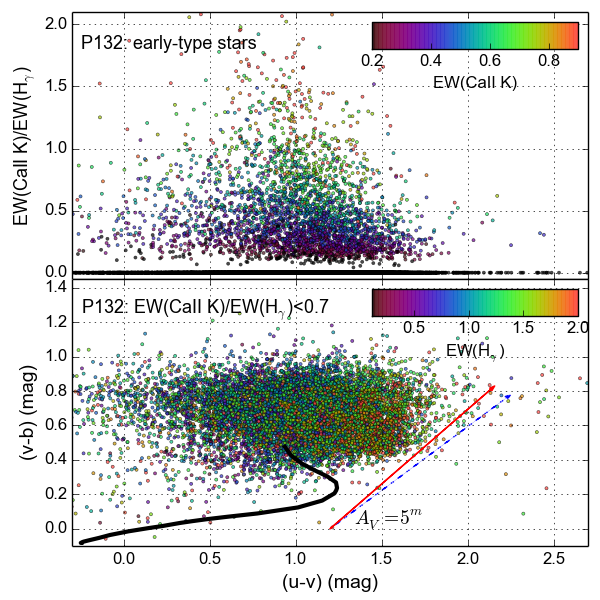}}
\caption[]{ Ratio of EW(\KCa)/EW(\Hgam) and (v-b) as a function of (u-v) for
  early-type sources detected on the P132 plates.  The thick line indicates the
  intrinsic colors of the stellar main-sequence.  The two reddening vectors
  used are shown for A$_\mathrm{V}$ = 5$^\mathrm{m}$. }
\label{fig:uvb}
\end{figure}

The (u-v)--(v-b) diagrams for early-type stars are shown in Fig.\ref{fig:uvb}
for the P132 plates.  The color relation for the intrinsic stellar "zero-age"
main-sequence (ZAMS) was taken from the models by \citet{marigo08} using a
1\,Myr isochrone with solar abundances. An isochrone with a metallicity Z=0.03
was also used but yielded similar results for the early-type stars with
effective temperatures in the range $3.8 < \log(\mathrm{T_{eff}})< 4.5$
(i.e. with EW(\Hgam) $>$ 0.2\,nm).  The maximum values of (u-v) in this
temperature range is 1.23 for $\log(\mathrm{T_{eff}}) = 3.96$.  The synthetic
spectra by \citet{munari05} show that stars with this temperature have a ratio
EW(\KCa)/EW(\Hgam) of 0.7 which was subsequently used to distinguish which
branch of the ZAMS to use for the estimate of extinctions.  The color excesses
for the indices are given by \citet{schlegel98} as E(v-b)/\AV\ = 0.152 and
E(u-v)/\AV\ = 0.205, while \citet{schlafly11} estimate the excesses to be
0.162 and 0.187, respectively.

The stars in the CCD show a distribution that has the same general slope as
the stellar ZAMS, but shifted by 1-4$^\mathrm{m}$ of visual extinction.  The
bluest part has, on average, higher extinctions which is consistent with it
containing intrinsic bright, early B-stars, which can be observed at larger
distances.  The color excesses by \citet{schlafly11} were used since they
yielded a larger number of stars, which is consistent with the intrinsic ZAMS.
The distribution of stars with EW(\KCa)/EW(\Hgam) $>$ 0.7 extends too far into
the blue, possibly because of misidentified lines in the late A- and F-star
spectra.

\begin{figure}
\resizebox{\hsize}{!}{\includegraphics{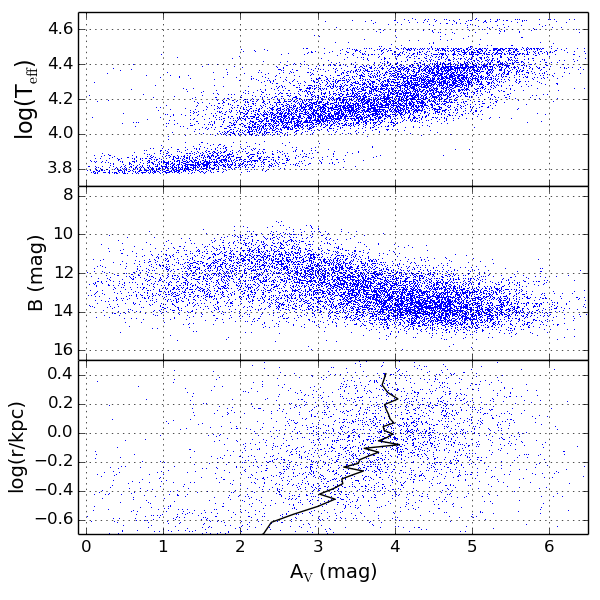}}
\caption[]{Distributions of \Teff, B, and distance r from the sun of
  early-type stars on the P132 plate set as a function of the visual
  extinction \AV\ estimated.  The fully drawn line in the lower diagram
    represents the average values of \AV.}
\label{fig:avd}
\end{figure}

Visual extinctions \AV\ were only estimated for stars with colors that are
consistent with them being reddened from the ZAMS, taking their EW(\KCa) into
consideration, which yields a total of 9\,571 stars earlier and 1\,504 stars
later than A1.  Their \Teff\ were computed using the intrinsic (u-v) color of
the isochrone.  For B-type stars, the strength of the Balmer lines indicates
the surface gravity g and can be used to correct absolute magnitude \MB\ for
the stellar evolution above the ZAMS.  The spectral models by \citet{munari05}
were used to estimate the relation between EW(\Hdel), \Teff, and
log(g). \MB\ was calculated by an interpolation in the ZAMS isochrone using
\Teff\ and adding a correction that was dependent on EW(\Hdel) for B-stars.
Finally, distances were calculated using \AB/\AV\ = 1.323.

The distribution of distances, B, and \Teff\ as functions of \AV,\ are given
in Fig.~\ref{fig:avd}.  In a magnitude limit sample, hot bright stars can be
observed at larger distances and higher extinctions than cooler stars, which
yield a general correlation between \AV\ and \Teff.  The deficiency of stars
near A0 is caused by the local maximum in (u-v) where photometric errors
spread sources to either side of the peak.  Sources with low extinction (i.e.,
\AV\ $< 2^\mathrm{m}$) are mostly cool, nearby stars and are evenly
distributed in apparent magnitude within the limits of the sample.  For higher
extinction, an anti-correlation between B and \AV\ is seen, which is a
combined effect of the large volume surveyed and the high extinction.

For a given distance, the scatter in \AV\ is very significant, which reflects
the patchy nature of the interstellar medium.  The average extinction
increases within the first kpc, while it is almost constant at greater
distances. The latter is caused by a selection effect since only bright stars
with relative low absorption can be observed.  The photometric uncertainties
in the (u-v) and (v-b) indices are $\sim$0\fm15, which translates to an error
$\sigma_\mathrm{A_V} = 0\fm3$.  Adding the uncertainty in B of
$\sigma_\mathrm{B}$ = 0\fm23, the distance modulus has an error
$\sigma_\mathrm{(m-M)} \approx 0\fm4, $ which corresponds to a relative error
in distance of 20--30\%.

%---------------------Section: Comparison with 2MASS-----------
\section{Comparison with the 2MASS survey}
\label{2mass}
The high extinction in the direction of the Galactic center makes it
interesting to cross-identify the sources found on the objective prism plates
with NIR surveys since they are less affected by dust attenuation.  The 2MASS
catalog \citep{2mass} and the VVV survey \citep{vvv} are the most appropriate
since their limiting magnitudes and area overlap match those of the current
survey. A significant issue is the high density of stars toward the Galactic
center, which makes an unambiguous identification difficult considering the
astrometric uncertainties of the spectra (i.e., up to 6\arcsec\ in $\delta$).
A more reliable identification can be obtained by matching spectra with Balmer
lines only to early-type stars in the NIR surveys.  Assuming the NIR color
excesses used by \citet{indebetouw05}, the reddening independent color index Q
= (H-K) $-$ 0.563*(J-H) was used to select early-type candidates with $-0\fm05
< \mathrm{Q} < 0\fm05$ corresponds to log(\Teff) $>$ 3.8 according to the
isochrone used above.

For the early-type stars on the P132 plates, 5\,350 (45\%) stars could be
matched with 2MASS sources within $\Delta(\alpha,\delta)$ = (3\arcsec,
16\arcsec) (i.e., three times their positional errors).  The closest match was
selected when multiple early-type candidates were found within the search
area, which happened for 31\% of the stars.  The positional differences of the
matches were $\alpha_\mathrm{ops} - \alpha_\mathrm{2MASS} =
0\farcs0\pm1\farcs0$ and $\delta_{ops} - \delta_{2MASS} =
-4\farcs5\pm5\farcs9$. A similar cross-identification was done with the VVV
catalog although its first data release did not cover the full area.  The VVV
catalog contains many faint non-stellar sources which are much redder than
stellar objects.  This type of wrong cross-identifications were reduced by
imposing a limit of K $>2.6\times$(J-K) + 14$^\mathrm{m}$ on matches.  A total
of 2 350 (35\%) matches were found with $\alpha_\mathrm{ops} -
\alpha_\mathrm{VVV} = -0\farcs0\pm1\farcs5$ and $\delta_{ops} - \delta_{VVV} =
-0\farcs2\pm8\farcs6$.  The large errors in $\delta$ were mainly located along
the edges of the scans which indicates problems in the extrapolation of the
astrometric solutions in this coordinate.  A similar issue was not observed in
$\alpha$.

\begin{figure}
\resizebox{\hsize}{!}{\includegraphics{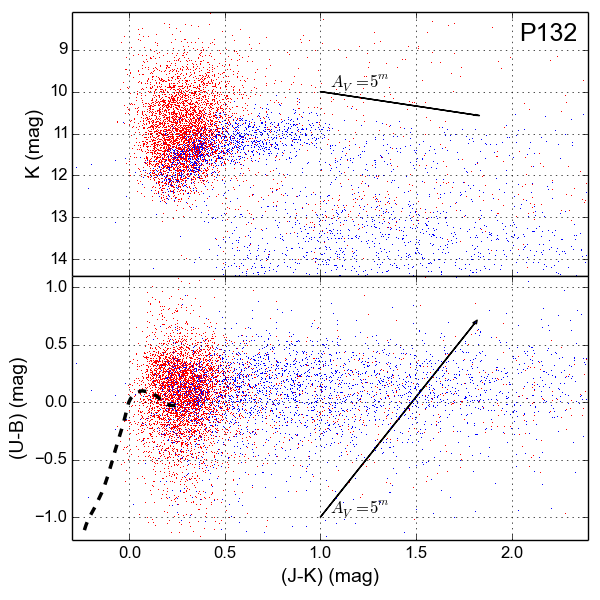}}
\caption[]{Color-magnitude and color-color diagram of early-type sources
  detected on the P132 plates and cross-identified with 2MASS (red) and VVV
  (blue). The top panel shows the (J-K)--K diagram while (J-K)--(U-B) is
  displayed at the bottom.  The main sequence for unreddened early-type
  stars is indicated by a black dashed line while the reddening vector is
  shown by an arrow.}
\label{fig:jhk}
\end{figure}

The (J-K)--K and (J-K)--(U-B) diagrams for early-type sources on P132 that
matched the NIR surveys are shown in Fig.~\ref{fig:jhk}, where average stellar
colors and reddening vector are also plotted.  The long tail of matches with
0\fm5 $<$ (J-K) and $12^\mathrm{m} <$ K are likely non-stellar sources
erroneously associated with the stellar spectra.  Most of these faint,
non-stellar objects were identified on the VVV survey, which is significantly
deeper than 2MASS.  Because of saturation, many early-type sources are not
matched with VVV.

In principle, extinctions can also be derived from the NIR data since
EW(\Hgam) can be used to avoid the confusion between reddened early-type stars
and nearby late-type stars.  Unfortunately, the color variation for early-type
stars is too close to the reddening vector to make this feasible.  Furthermore,
wrong identifications yield a significant scatter.

%--------------------- Section: Distribution of Early-types -------------------
\section{Properties of catalog}
\label{distribution}
\begin{figure}
\resizebox{\hsize}{!}{\includegraphics{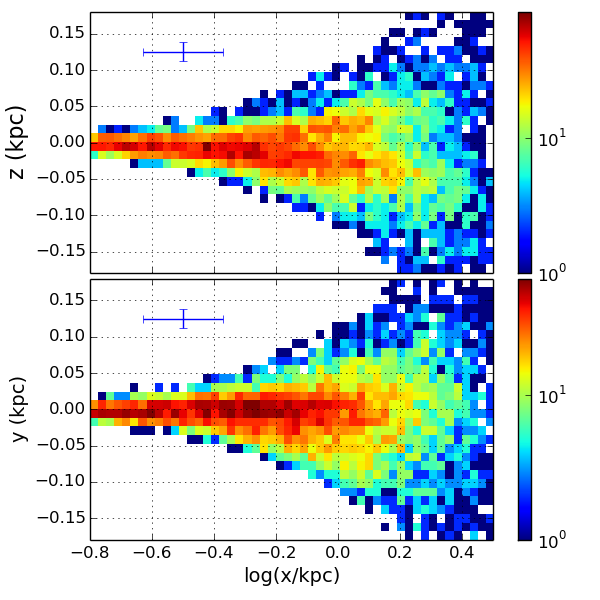}}
\caption[]{ Histograms of the spatial distribution of early-type stars on the
  P132 plates.  The top plot shows the distribution perpendicular to the plane
  toward the Galactic center while the bottom diagram displays the histogram
  of stars in the Galactic plane. A typical error bar is shown for a distance
  error of 30\% at \lII\ = \bII\ = 2\fdg5. }
\label{fig:rbd}
\end{figure}

The distributions of the early-type stars on the P132 plates are shown in
Fig.~\ref{fig:rbd} as histograms that use a right-hand Cartesian coordinate
system centered on the Sun, and with the x-axis pointed toward the Galactic
center.  A typical error bar is indicated on the figure corresponding to a
relative error of 30\% in distance at \lII\ = \bII\ = 2\fdg5.  The projection
on the plane shows a smooth distribution, which is expected for a
magnitude-limited sample, taking the survey area into consideration.  More
structure is seen perpendicular to the plane where a deficiency of stars in
the plane is seen for distances larger than 0.5\,kpc, likely associated with
an increase in extinction in the Sagittarius arm.

\begin{figure}
\resizebox{\hsize}{!}{\includegraphics{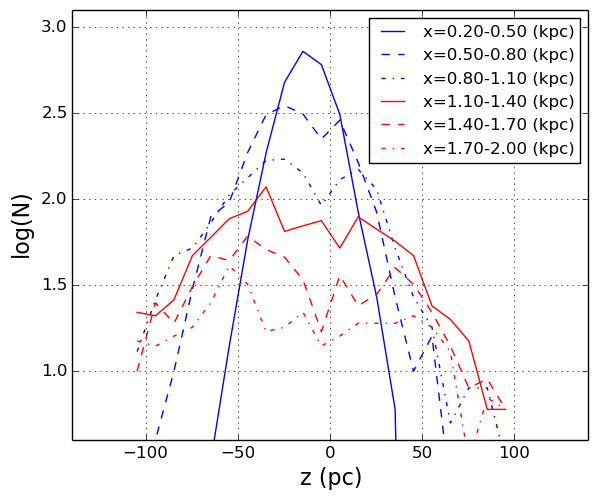}}
\caption[]{Distribution of early-type stars on the P132 plates within a
  $\pm$3\degr\ cone around the direction to the Galactic center as a function
  of the height above the plane, $z$, and distance from the sun, $x$.}
\label{fig:rzd}
\end{figure}

The stellar densities of early-type stars in a cone of $\pm$3\degr\ around the
Galactic center direction were computed as a function of $x$ and $z$ (see
Fig~\ref{fig:rzd}).  Cells of 10\,pc were used perpendicular to the plane
while a linear binning of 0.3\,kpc was applied in the direction toward the
Galactic center.  The closest bin (i.e., $x$=0.2-0.5\,kpc) shows an almost
symmetric distribution with a centroid below the plane.  All the more distant
radial bins display significant dips (i.e., $>5\sigma$) close to the plane
associated with the extinction in the Sagittarius arm.  All radial bins have a
centroid below the plane that ranges from -10\,pc for x=0.5\,kpc to -22\,pc at
a distance of 1.5\,kpc, which is consistent with \citet{joshi07}.

%--------------------- Section: Conclusion ---------------------
\section{Discussion and conclusion}
\label{discussion}
Of the three plate sets available, the P132 plates with IIIa-J emulsions
turned out to be the best for detecting faint early-type stars in the
direction of the Galactic center.  More sources were found on the P131 films
but most of them were red, late-type stars because of the high sensitivity in
the red of the 4415 emulsions.  Furthermore, the long spectra on the P131
films made crowding a substantial issue.  The old IIa-O were significantly
shallower than the other two plate sets, because of both the emulsion and the
widened spectra.

The spectra provide both general colors and EWs of the most prominent lines,
which allow a reliable selection of early-type stars on the plates with 12675
candidates identified on the P132 plate set.  More importantly, the
comparison of the EWs of Balmer lines and \KCa\ enables a separation of stars
that are hotter and cooler than A0 and, consequently, a resolution of the
ambiguity in the estimate of individual extinctions using CCDs.

Although variable extinction and crowding make it impossible to evaluate a
reliable volume density of stars as a function of distance, the survey area is
large enough to obtain a reasonable sample of early-type stars toward the
Galactic center.  Assuming a smooth, intrinsic distribution function of the
early-type stars, one can identify general features in the structure of the
extinction.  The most prominent is the increase of absorption in the Galactic
plane which starts at a distance of 0.5-0.8\,kpc from the sun and is
associated with the Sagittarius arm, which is located at a Galactic centric
distance of 6.6\,kpc \citep{reid14}.

The increase of extinction in the plane suggests the presence of star
formation associated with the Sagittarius spiral arm but does not indicate
that a mass perturbation is associated with it.  The safest way to decide if
it is a major arm or not is to measure the velocity perturbation that is
induced by the arm.  The current survey of early-type stars has identified
more than 9\,571 candidates earlier than A1 in the direction of the Galactic
center at distances reaching beyond the Sagittarius arm.  This sample can be
used to study the kinematics of these stars as a function of distance and,
thereby, to firmly determine the mass perturbation of the arm.

%--------------------- Acknowledgements ------------------------
\begin{acknowledgements}
The main parts of the reductions were done with the ESO-MIDAS system while the
analysis employed Python scripts that utilize the {\it scipy} library.  This
research made use of the SIMBAD database, operated at CDS, Strasbourg,
France. This publication makes use of data products from the Two Micron All
Sky Survey, which is a joint project of the University of Massachusetts and
the Infrared Processing and Analysis Center/California Institute of
Technology, funded by the National Aeronautics and Space Administration (NASA)
and the National Science Foundation.  Helpful comments by an anonymous referee
were also appreciated.

\end{acknowledgements}
%--------------------- References ------------------------------
\bibliographystyle{aa}
\bibliography{AstronRef}

\begin{thebibliography}{29}
\expandafter\ifx\csname natexlab\endcsname\relax\def\natexlab#1{#1}\fi

\bibitem[{Bertin \& Arnouts(1996)}]{bertin96}
Bertin, E. \& Arnouts, S. 1996, A\&AS, 117, 393

\bibitem[{Dehnen \& Binney(1998)}]{dehnen98}
Dehnen, W. \& Binney, J. 1998, MNRAS, 298, 387

\bibitem[{Drimmel(2000)}]{drimmel00}
Drimmel, R. 2000, A\&A, 358, L13

\bibitem[{Englmaier \& Gerhard(1999)}]{englmaier99}
Englmaier, P. \& Gerhard, O. 1999, MNRAS, 304, 512

\bibitem[{Georgelin \& Georgelin(1976)}]{georgelin76}
Georgelin, Y.~M. \& Georgelin, Y.~P. 1976, A\&A, 49, 57

\bibitem[{Hauck \& Mermilliod(1998)}]{hauck98}
Hauck, B. \& Mermilliod, M. 1998, A\&AS, 129, 431

\bibitem[{H{\o}g {et~al.}(2000)H{\o}g, Fabricius, Makarov, Urban, Corbin,
  Wycoff, {et~al.}}]{hog00}
H{\o}g, E., Fabricius, C., Makarov, V.~V., {et~al.} 2000, A\&A, 355, L27

\bibitem[{Indebetouw {et~al.}(2005)Indebetouw, Mathis, Babler, Meade, Watson,
  Whitney, {et~al.}}]{indebetouw05}
Indebetouw, R., Mathis, J.~S., Babler, B.~L., {et~al.} 2005, ApJ, 619, 931

\bibitem[{Joshi(2007)}]{joshi07}
Joshi, Y.~C. 2007, MNRAS, 378, 768

\bibitem[{Lin \& Shu(1964)}]{lin64}
Lin, C.~C. \& Shu, F.~H. 1964, ApJ, 140, 646

\bibitem[{Lin {et~al.}(1969)Lin, Yuan, \& Shu}]{lin69}
Lin, C.~C., Yuan, C., \& Shu, F.~H. 1969, ApJ, 155, 721

\bibitem[{Marigo {et~al.}(2008)Marigo, Girardi, Bressan, Groenewegen, Silva, \&
  Granato}]{marigo08}
Marigo, P., Girardi, L., Bressan, A., {et~al.} 2008, A\&A, 482, 883

\bibitem[{Mermilliod(1991)}]{mermilliod91}
Mermilliod, J.~C. 1991, {Catalogue of Homogeneous Means in the UBV System},
  Tech. rep., Institut d'Astronomie, Universite de Lausanne

\bibitem[{Mongui{\'o} {et~al.}(2015)Mongui{\'o}, Grosb{\o}l, \&
  Figueras}]{monguio15}
Mongui{\'o}, M., Grosb{\o}l, P., \& Figueras, F. 2015, A\&A, 577, A142

\bibitem[{Munari {et~al.}(2005)Munari, Sordo, Castelli, \& Zwitter}]{munari05}
Munari, U., Sordo, R., Castelli, F., \& Zwitter, T. 2005, A\&A, 442, 1127

\bibitem[{Nonino {et~al.}(1999)Nonino, Bertin, da~Costa, Deul, Erben, Olsen,
  {et~al.}}]{nonino99}
Nonino, M., Bertin, E., da~Costa, L., {et~al.} 1999, A\&AS, 137, 51

\bibitem[{Payne \& Williams(1929)}]{payne29}
Payne, C. \& Williams, E. T.~R. 1929, MNRAS, 89, 526

\bibitem[{Reid {et~al.}(2014)Reid, Menten, Brunthaler, Zheng, Dame, Xu,
  {et~al.}}]{reid14}
Reid, M.~J., Menten, K.~M., Brunthaler, A., {et~al.} 2014, ApJ, 783, 130

\bibitem[{Russeil {et~al.}(2007)Russeil, Adami, \& Georgelin}]{russeil07}
Russeil, D., Adami, C., \& Georgelin, Y.~M. 2007, A\&A, 470, 161

\bibitem[{Saito {et~al.}(2012{\natexlab{a}})Saito, Hempel, Minniti, Lucas,
  Rejkuba, Toledo, {et~al.}}]{vvv}
Saito, R.~K., Hempel, M., Minniti, D., {et~al.} 2012{\natexlab{a}}, A\&A, 537,
  A107

\bibitem[{Saito {et~al.}(2012{\natexlab{b}})Saito, Minniti, Dias, Hempel,
  Rejkuba, Alonso-Garc{\'i}a, {et~al.}}]{saito12}
Saito, R.~K., Minniti, D., Dias, B., {et~al.} 2012{\natexlab{b}}, A\&A, 544,
  A147

\bibitem[{Sampson(1925)}]{sampson25}
Sampson, R.~A. 1925, MNRAS, 85, 212

\bibitem[{Schlafly \& Finkbeiner(2011)}]{schlafly11}
Schlafly, E.~F. \& Finkbeiner, D.~P. 2011, ApJ, 737, 103

\bibitem[{Schlegel {et~al.}(1998)Schlegel, Finbeiner, \& Davis}]{schlegel98}
Schlegel, D.~J., Finbeiner, D.~P., \& Davis, A. 1998, ApJ, 500, 525

\bibitem[{Skrutskie {et~al.}(2006)Skrutskie, Cutri, Stiening, Weinberg,
  Schneider, Carpenter, {et~al.}}]{2mass}
Skrutskie, M.~F., Cutri, R.~M., Stiening, R., {et~al.} 2006, AJ, 131, 1163

\bibitem[{Str{\"o}mgren(1966)}]{stromgren66}
Str{\"o}mgren, B. 1966, ARA\&A, 4, 433

\bibitem[{Vall{\'e}e(2014)}]{vallee14}
Vall{\'e}e, J.~P. 2014, AJ, 148, 5

\bibitem[{Wielen(1977)}]{wielen77}
Wielen, R. 1977, A\&A, 60, 263

\bibitem[{Y{\'a}{\~n}ez {et~al.}(2008)Y{\'a}{\~n}ez, Norman, Martos, \&
  Hayes}]{yanez08}
Y{\'a}{\~n}ez, M.~A., Norman, M.~L., Martos, M.~A., \& Hayes, J.~C. 2008, ApJ,
  672, 207

\end{thebibliography}
%--------------------- Apendix ------------------------------
\Online
\begin{appendix} 
\section{Protometric transformations}
The photometric transformations from the standard system to observed
magnitudes were estimated by linear least-square fits of the form
$$ \mathrm{X_{obs}} = \mathrm{c_X} + \mathrm{a_X} \mathrm{X_{std}} \pm
\sigma_\mathrm{x,}$$ where X$_\mathrm{std}$ denotes the standard magnitude or
color index, while X$_\mathrm{obs}$ is the observed values.  The constant
c$_\mathrm{X}$ and the linear term a$_\mathrm{X}$ are listed in
Table~\ref{tab:trans}, where the root-mean-square of the fit
$\sigma_\mathrm{X}$ and the number of sources used n$_\mathrm{X}$ are also
given.

\begin{table*}
\caption[]{Photometric transformations. }
\label{tab:trans}
\begin{tabular}{l rr rr rr rr rr rr}
\hline\hline
 Plates &
 \multicolumn{1}{c}{c$_\mathrm{B}$} & \multicolumn{1}{c}{a$_\mathrm{B}$} & 
 \multicolumn{1}{c}{c$_\mathrm{(U-B)}$} & 
 \multicolumn{1}{c}{a$_\mathrm{(U-B)}$} &
 \multicolumn{1}{c}{c$_\mathrm{(B-V)}$} & 
 \multicolumn{1}{c}{a$_\mathrm{(B-V)}$} &
 \multicolumn{1}{c}{c$_\mathrm{(u-b)}$} & 
 \multicolumn{1}{c}{a$_\mathrm{(u-b)}$} &
 \multicolumn{1}{c}{c$_\mathrm{c_1}$} & 
 \multicolumn{1}{c}{a$_\mathrm{c_1}$} &
 \multicolumn{1}{c}{c$_\mathrm{H_\beta}$} &
 \multicolumn{1}{c}{a$_\mathrm{H_\beta}$} \\
 & \multicolumn{1}{c}{$\sigma_\mathrm{B}$} &
 \multicolumn{1}{c}{n$_\mathrm{B}$} &
 \multicolumn{1}{c}{$\sigma_\mathrm{(U-B)}$} &
 \multicolumn{1}{c}{n$_\mathrm{(U-B)}$} &
 \multicolumn{1}{c}{$\sigma_\mathrm{(B-V)}$} & 
 \multicolumn{1}{c}{n$_\mathrm{(B-V)}$} &
 \multicolumn{1}{c}{$\sigma_\mathrm{(u-b)}$} & 
 \multicolumn{1}{c}{n$_\mathrm{(u-b)}$} &
 \multicolumn{1}{c}{$\sigma_\mathrm{c_1}$} &
 \multicolumn{1}{c}{n$_\mathrm{c_1}$} &
 \multicolumn{1}{c}{$\sigma_\mathrm{H_\beta}$} &
 \multicolumn{1}{c}{n$_\mathrm{H_\beta}$} \\ \hline 
 P023 & 
       -15.469 & 0.808 &   1.687 & 0.816 &       - &     - &
        -0.406 & 1.016 &  -0.020 & 0.622 &      - &     - \\
   &    0.173 &  296 &  0.132 &  261 &      - &     - & 
        0.174 &   25 &  0.068 &   25 &      - &     - \\
 P131 &
       -16.428 & 0.918 &   1.529 & 0.753 &   0.096 & 1.297  &
        -0.737 & 1.027 &   0.397 & 1.013 &  -0.727 & 1.026 \\
   &    0.252 &  285 &  0.178 &  258 &  0.236 &  290 &
        0.151 &   13 &  0.101 &   13 &  0.029 &    4  \\
 P132 &
        -16.540 & 0.843 &   1.362 & 1.247 &  -1.071 & 0.812 &
         -1.719 & 1.315 &   0.202 & 1.244  &     - &     - \\
   &    0.234 & 136 &  0.175 & 122 &  0.243 & 136 & 
        0.076 &   7 &  0.119 &  15 &      - &     - \\
 P13232 &
        -16.151 & 0.632 &   1.473 & 0.705 &   0.424 & 0.740 &
         -0.578 & 0.823 &   0.257 & 1.387 &      - &     - \\
   &    0.344 &  10 &  0.141 &  8 &  0.189 &  10 & 
        0.223 &   4 &  0.141 &  4 &      - &     - \\  \hline
\end{tabular}
\end{table*}

\section{Data table}
\label{app:data}
The list of early-type stars on the P132 plate set is available through
CDS\footnote{Centre de Donn{\'e}es astronomiques de Strasbourg:
  http://cds.u-strasbg.fr} as a FITS table with the columns listed in
Table~\ref{tab:data}.  Besides identifiers and coordinates, the synthetic
photometry and equivalent width of main lines are provided.  The original
spectra scanned on the photographic plates are included in relative intensity
units.  The derived values for the individual extinctions and distances are
also given.  Tables for the other plate sets are available from the author.

\begin{table}
\caption[]{Column specifications for the table with data of early-type stars
  from P132 plate set.}
\label{tab:data}
\begin{tabular}{llll}
 \hline\hline
 Label & Format & Unit & Remarks \\ \hline
id\_ops   & A17   &   - & Identifier \\
RAdeg     & D     & deg & Right Ascension in J2000 \\
DECdeg    & D     & deg & Declination  in J2000  \\
id\_scan  & A10   & -   & Scan identifier \\
Bmag      & E     & mag & Johnson B magnitude \\
U\_Bj     & E     & mag & Johnson (U-B) index \\
u\_vs     & E     & mag & Str\"{o}mgren (u-v) index \\
v\_bs     & E     & mag & Str\"{o}mgren (v-b) index \\
Hg\_ew    & E     & nm  & Equivalent width of \Hgam \\
Hd\_ew    & E     & nm  & Equivalent width of \Hdel \\
He\_ew    & E     & nm  & Equivalent width of \Heps \\
KCa\_ew   & E     & nm  & Equivalent width of \KCa \\
Av        & E     & mag & Extinction in V \\
logTe     & E     & -   & Logarithm of effective temperature \\
dBmag     & E     & mag & B magnitude above ZAMS \\
BMmag     & E     & mag & Absolute magnitude in Johnson B filter \\
dist      & E     & kpc & Distance from sun \\
start     & E     & um & Start of spectral scan relative to \Hgam \\
step      & E     & um & Step size of spectral scan \\
spec      & 1000E &    & Spectrum in relative intensities \\
\end{tabular}
\end{table}

\end{appendix}
\end{document}